\pdfoutput=1

\documentclass[twocolumn,a4paper]{article}

\usepackage{graphicx}

\begin{document}
\title{Mode shaping in mixed ion crystals of $^{40}$Ca$^{2+}$ and $^{40}$Ca$^+$ }

\author{T. Feldker,$^{1\ast}$ L. Pelzer,$^{1}$ M. Stappel,$^{1,2}$ P. Bachor,$^{1,2}$ R. Steinborn,$^{1,2}$ D. Kolbe,$^{1,2,3}$\\ 
J. Walz,$^{1,2}$ F. Schmidt-Kaler$^{1}$}

\maketitle

\normalsize{$^1$QUANTUM, Institut f\"ur Physik, Universit\"at Mainz, Staudingerweg 7, D-55128 Mainz, Germany}\\
\normalsize{$^2$Helmholtz-Institut Mainz, D-55099, Germany}\\
\normalsize{$^3$\emph{Present address:} Deutsches Zentrum f\"ur Luft- und Raumfahrt e. V., Institut f\"ur Technische Physik, Pfaffenwaldring 38-40, D-70569 Stuttgart, Germany}

\begin{abstract}
We present studies of mixed Coulomb crystals of $^{40}$Ca$^+$ and $^{40}$Ca$^{2+}$ ions in a linear Paul trap. Doubly charged ions are produced by photoionisation of trapped $^{40}$Ca$^+$ with a vacuum ultraviolet laser source and sympathetically cooled via Doppler cooled $^{40}$Ca$^+$ ions. We investigate experimentally and theoretically the structural configurations and the vibrational modes of these mixed crystals. Our results with $^{40}$Ca$^{2+}$ are an important step towards experimental realization of the proposals for mode shaping in a linear crystal and spin-dependent configuration changes from zigzag to linear as proposed by Li {\em{et al.}}, Phys. Rev. A 87, (2013) 052304 and Phys. Rev. Lett 108, (2012) 023003 using ions excited to Rydberg states.
\end{abstract}
\section{Introduction}
\label{intro}

Trapped ions are among the most promising systems for the realization of a quantum computer. While excellent control over single qubits\,\cite{Blatt08}, high fidelity entangling of two qubits\,\cite{benhelm08,Home09}, simple quantum algorithms\,\cite{Ri04,La11} and entangling of up to 14 qubits\,\cite{Mo11} have been demonstrated, the scaling to large numbers of qubits remains challenging due to the increasingly complex mode spectrum of large ion crystals. 

Performing high fidelity quantum gate operations even in long linear ion crystals\,\cite{Lin09} using optimized laser pulses\,\cite{Ga03,Du04} is an highly interesting yet challenging approach to scale up the number of qubits available for quantum information processing with trapped ions. The idea of mode shaping has the potential to drastically simplify this approach since the complexity of the mode spectrum can be significantly reduced. Such tailoring of vibrational modes may be possible employing the unique properties of Rydberg states\,\cite{Li12}, where the polarizability of the ion is strongly enhanced leading to significant modifications of the radial trapping potential\,\cite{Schmidt-Kaler11,Mueller08}. 

Equally fascinating is the perspective of spin-dependent trap-ion interactions exhibiting a strength rivalling the Coulomb forces, such that structural configuration changes of ion crystals are induced. Again, Rydberg excitations have been proposed to induce such a phase transitions\,\cite{Li12a} between a linear string and a zigzag crystal. Hence, Rydberg excitation of ions may be used to quench an ion crystal\,\cite{Ulm13,Pyka13}, producing a non-equilibrium situation and eventually even enter the quantum regime\,\cite{Ba12} of this phase transition.

\begin{figure*}
	\centering
		\includegraphics[width=1\textwidth]{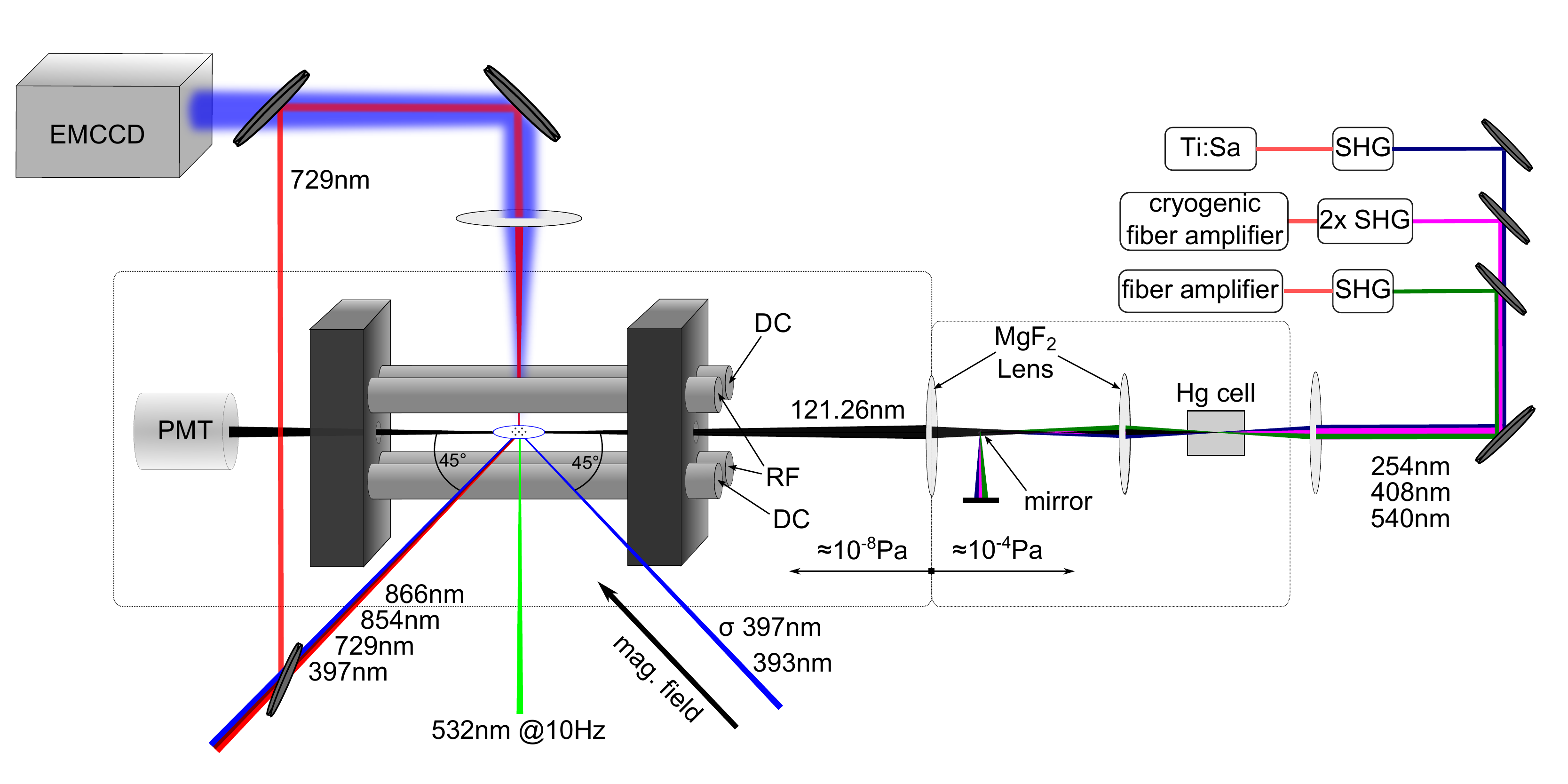}
\caption{Experimental setup showing the Paul trap\, with laserbeams for trapping, cooling and spectroscopy of $^{40}$Ca$^+$ and the EMCCD detection system. The vacuum ultraviolet (VUV) generation is accomplished by four-wave-mixing (FWM) in a Hg-vapour cell. The fundamental beams are frequency doubled by second-harmonic generation (SHG), overlapped and focussed into the Hg-vapour. After the FWM region they are dispersively separated from the VUV light with a MgF$_2$ lens (f= 130\,mm) and a tiny mirror. A second MgF$_2$ lens (f= 125\,mm) at a distance to the ions of about 240\,mm, focusses and alignes the VUV beam to the ion crystal and is used for vacuum separation between the ion trap and the FWM region.}
\label{fig:setup}       
\end{figure*}

The experimental work presented here features important aspects of mode shaping using mixed crystals with doubly charged ions instead of Rydberg excitations. We investigate in detail radial modes and configuration changes, thus providing a complete understanding of the mode shaping which would be required for quantum gate operation. For the measurements we benefit from the effectively infinite lifetime of  $^{40}$Ca$^{2+}$ ions as compared to a lifetime of about 100\,$\mu$s for a Rydberg excitation. The presented experiments with $^{40}$Ca$^{2+}$ ions can be considered as an important step towards mode shaping experiments with Rydberg excitations. Still, one important aspect of the proposed mode shaping and ion crystal configuration is not captured: Only coherent Rydberg excitation would give complete control over these effects, including superposition states. 
   
We generate mixed Coulomb crystals consisting of ions with a large difference in charge-to-mass ratio by ionizing single $^{40}$Ca$^+$ ions with a vacuum ultraviolet (VUV) source\,\cite{Kolbe12}. The wavelength of the VUV source can be changed from ionisation to Rydberg excitation wavelength and a very similar excitation scheme is used in both cases.

The paper is organized as follows: After describing the experimental setup we introduce the photoionisation scheme for selectively ionizing trapped $^{40}$Ca$^+$. In the following we describe the frequency measurement of various modes of mixed crystals and compare the results to simulations. We also investigate structural changes of the crystals, as depending on the trap parameters and the position of the $^{40}$Ca$^{2+}$ ion. In a last section we investigate local modes created by a $^{40}$Ca$^{2+}$ impurity in a longer ion chain.

\section{Experimental setup}
\label{expsetup}
The ions are trapped in a linear Paul trap (Fig.\ \ref{fig:setup}, described in detail in\,\cite{Schmidt-Kaler11}) consisting of four cylindrical electrodes (diameter 2.5\,mm) and two endcaps (distance 10\,mm) with holes (diameter 1\,mm) to provide axial optical access to the ions. The Paul trap is operated with an rf-amplitude of $U_{\textnormal{rf}}$ = 100\,V - 200\,V at a frequency $\Omega$/(2$\pi$) = 10.66\,MHz. An additional offset voltage of $U_{\textnormal{of}}$ = -1.5\,V is applied to the DC-electrodes to lift the radial degeneracy of the trapping potential. The resulting trapping frequencies are $\omega_x$/(2$\pi$) = 150\,kHz - 500\,kHz and $\omega_y$/(2$\pi$)= 450\,kHz - 650\,kHz depending on the rf-amplitude. The endcaps are operated at a DC voltage of 400\,V which yields an axial trapping frequency of $\omega_z$/(2$\pi$) = 119\,kHz. 

\begin{figure}
	\centering
		\includegraphics[width=0.5\textwidth]{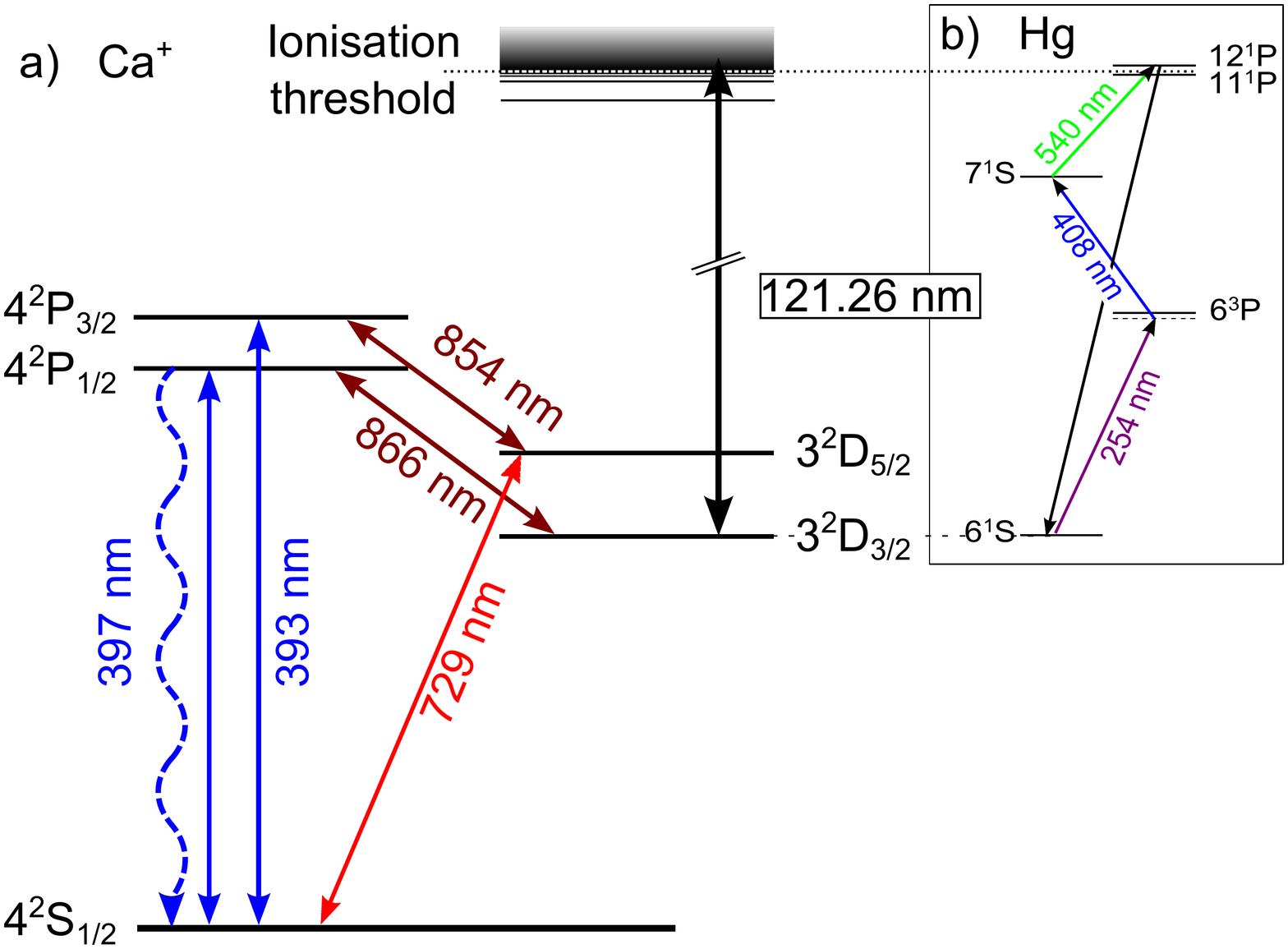}
\caption{a) Level scheme of $^{40}$Ca$^{+}$ with transitions at 397\,nm used for Doppler cooling 
and detection, repumping at 866\,nm and 854\,nm, spectroscopy at 729\,nm and ionisation at 121.26\,nm.
The laser near 393\,nm allows for pumping to the $3^2$D$_{5/2}$ state. b) Relevant Hg levels used in the four wave mixing process. The UV and blue lasers are fixed to the two-photon resonance $6^1\textnormal{S}\rightarrow7^1\textnormal{S}$, the VUV wavelength is determined by the third laser.}
\label{fig:excitation}       
\end{figure}

 The ions are produced by photoionisation with a pulsed laser at $532$\,nm from a neutral atom beam of $^{40}$Ca. Ions are cooled and repumped by copropagating laserbeams at wavelengths of $397$\,nm, $866$\,nm and $854$\,nm. With a diode laser at $393$\,nm ions can be optically pumped into the $3^2$D$_{5/2}$ state with nearly 100\% efficiency. The level scheme of $^{40}$Ca$^+$ ions including important transitions and laser wavelenghts is depicted in Fig.\ \ref{fig:excitation}. All laser-beams are switched by acousto-optic modulators (AOM) in double pass configuration. Additionally, a stabilized ($\Delta \nu <$ 1~kHz) Titanium:Sapphire (Ti:Sa) laser at $729$\,nm allows for resolving motional sidebands on the quadrupole transition $4^2$S$_{1/2} \rightarrow 3^2$D$_{5/2}$. As depicted in Fig.\ \ref{fig:setup}, there are two spectroscopy beams available, one overlapped with the Doppler cooling beam, having a projection on all principle trap axes and one propagating through the high numerical aperture (NA) camera lens. The latter is tightly focussed to about $5\,\mu$m and can be used for addressing single ions in a bigger crystal.

 The imaging system consists of a lens with focal length 66.8\,mm which maps the ion crystal via a folding mirror on an EMCCD (Electron Multiplying Charge Coupled Device) chip. The numerical aperture of 0.27 limits the theoretical resolution to 0.9 $\mu$m, the magnification can be determined by comparison of simulated and measured ion positions to a value of 17. The exposure time for a single picture may be varied between 1 and 50\,ms. 
 
 For single-photon ionisation a continuous wave coherent light source at 121.26\,nm is used. A detailed description of this laser can be found in\,\cite{Kolbe12,Schmidt-Kaler11}. Coherent VUV radiation at 121.26\,nm is produced by triple resonant four-wave mixing in mercury with three fundamental waves at 254\,nm, 408\,nm and 540\,nm (see Fig.\ \ref{fig:excitation}b). In the four wave mixing scheme the three fundamental lasers are tuned close to the transitions $6^{1}$S$-6^{3}$P\,\cite{Ko12}, $6^{3}$P$-7^{1}$S and $7^{1}$S$-12^{1}$P in mercury, respectively, which significantly increases the yield of VUV power to about 1\,$\mu$W at 121.26\,nm. 
 The setup of the VUV laser is shown in Fig.\ \ref{fig:setup}. The three fundmental beams are superimposed by dichroic mirrors and focussed into mercury vapour for VUV generation. The fundamental beams are dispersively separated from the VUV light with a MgF$_2$ lens and a tiny mirror. With a second MgF$_2$ lens mounted on a 3D-translation stage the VUV light is focussed and aligned to the trap. The VUV radiation propagates through holes in the endcaps of the trap and is detected with a photo multiplier tube (PMT) behind the trap. The measured focus has a width of $\omega_0 \approx 20\,\mu$m.

For photoionisation of $^{40}$Ca$^{+}$ we first load a linear string of three ions. The ions are pumped to the $3^2$D$_{3/2}$ state by laser light at 397\,nm wavelength, before we apply ionization light near 121.26\,nm. Since a doubly-charged ion does not resonantly scatter light near 397\,nm a successful formation of $^{40}$Ca$^{2+}$ is characterized as a dark gap in the ion crystal. However, such dark gaps may also result from loading of ions from ionized background gas. In order to verify the ionization of a $^{40}$Ca$^{+}$ and to discriminate $^{40}$Ca$^{2+}$ from accidentally loaded ions of other species we check the length of the crystal, which increases by 21\% ($\approx 8\,\mu$m) for a $^{40}$Ca$^{2+}$ ion in central position. To produce larger crystals with an impurity ion we also start from a three ion crystal, ionize one of the ions and finally load the desired number of additional $^{40}$Ca$^+$ ions into the trap. This way we make sure that the crystal contains a $^{40}$Ca$^{2+}$ ion.

\section{Vibrational modes of mixed three-ion crystals}
\label{modes3}

The $^{40}$Ca$^{2+}$ impurity leads to modificated vibrational modes compared to a $^{40}$Ca$^{+}$ crystal. The broken symmetry of the crystal gives rise to radial modes which, except for the rocking mode with $^{40}$Ca$^{2+}$ in central position\,($\uparrow\circ\downarrow$), are no longer orthogonal to the center of mass mode. Hence, modes can be excited by applying a resonant radiofrequency to one of the DC-electrodes. We concentrate on the radial modes of the crystal since these are the most interesting for mode shaping. The axial modes of a three-ion crystal with $^{40}$Ca$^{2+}$ have been measured previously by Kwapien et al.\,\cite{Kw07}. 

\begin{figure}
	\centering
		\includegraphics[width=0.5\textwidth]{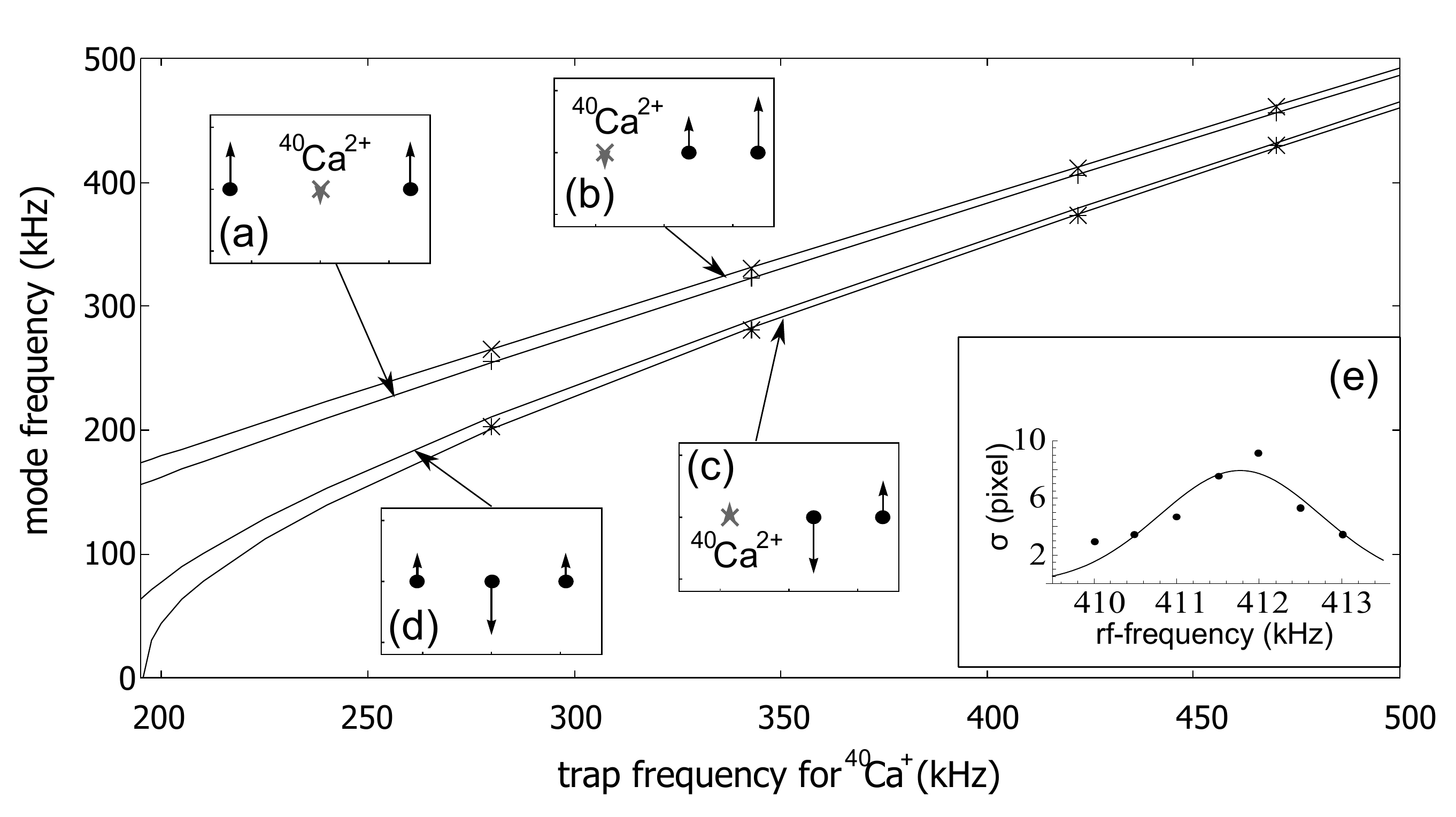} 
	\caption{Measured and simulated frequencies of selected modes of mixed ion crystals. Dots represent measured values while lines are from numerical simulations. The small insets (a-c) show the modes which where measured. For comparison the simulated frequency of the zigzag mode (d) without $^{40}$Ca$^{2+}$ is also shown. (e) The oscillation amplitude of the ions is plotted for different radiofrequencies. The eigenfrequency is determind as the center frequency of a gaussian fitted to the oscillation amplitude as a function of the applied rf-frequency.}
	\label{fig:ModePlot}
\end{figure}

For frequeny determination the oscillation amplitude of the ions with respect to the applied radiofrequency at a constant rf-amplitude is measured. The oscillation amplitude results in a smeard out fluorescence of the ion and can be determined from the camera image. The eigenfrequency of the mode is determined as the central frequency of a gaussian, fitted to the oscillation amplitude as a function of the applied rf-frequency. We avoid non-linear response\,\cite{Akerman10} by keeping a low excitation amplitude. The simulated and measured frequencies are shown in Fig.\ \ref{fig:ModePlot} and agree to about 1\,kHz, the deviation can be explained by errors in the determination of the eigenfrequeny (0.5\,kHz) and a drift of the trapping potential between measurement and calibration where the trap frequencies $\omega_{x,y,z}$ are measured with a single $^{40}$Ca$^{+}$. The rf-amplitude $U_{\textnormal{rf}}$ is not completely constant but decreases slowly over time which leads to the observed trapping potential drift of $\Delta\omega_{x,y}/(2\pi)\approx$ 10\,kHz over the course of a day.  The highest frequency modes where the $^{40}$Ca$^{2+}$ exhibits an approximately 50 times higher oscillation amplitude than the $^{40}$Ca$^{+}$ are missing in the plot because they could not be measured exactly. This is because the $^{40}$Ca$^{2+}$ ion can not be detected while the oscillation amplitude of $^{40}$Ca$^{+}$ ions remains small.

The zigzag mode of a crystal with $^{40}$Ca$^{2+}$ in outer position (Fig.\,\ref{fig:ModePlot}c) is shifted to lower frequencies compared to the zigzag mode of a three-ion $^{40}$Ca$^{+}$ crystal, while the zigzag mode of a crystal with $^{40}$Ca$^{2+}$ in central position (Fig.\,\ref{fig:ModePlot}a) is shifted to higher frequencies. This result agrees with the observed behaviour of the zigzag transition in section\,\ref{conf}.

\section{Structural changes in a three-ion crystal by an $^{40}$Ca$^{2+}$ impurity}
\label{conf}

Assuming the well justified pseudo potential approximation\,\cite{Leibfried03}, a linear Paul trap is fully described by a static potential in axial direction with angular frequency $\omega_z = 2\sqrt{q\beta/M}$ and a pseudopotential in radial direction with angular frequency $\omega_r = \sqrt{2((q\gamma/(M\Omega))^2-q\beta/M)}$ where $q$ denotes the charge, $M$ the mass of the ion and $\gamma, \beta$ the curvatures of the electric potentials. By applying an offset voltage $V_{\textnormal{of}}$ to the DC-electrodes an additional static electric potential $\beta_\textnormal{rad}$ is induced and the radial pseudopotential reads:
\begin{equation}\label{eq:omegay}
	\omega_{x,y} = \sqrt{2((q\gamma/(M\Omega))^2 - q\beta/M \mp q\beta_{rad}/M)}
\end{equation}
The configuration of an ion crystal in this potential is determined by the anisotropy parameters $\alpha_{x,y}= \omega_z^2/ \omega_{x,y}^2$. For tight radial confinement $\omega_{x,y} \gg \omega_{z}$ the ions form a linear string aligned on the $z$-axis. Lowering the radial confinement causes a phase transition to a zigzag configuration at $\alpha=\alpha_\textnormal{crit}$\,\cite{Enzer00,Ulm13,Pyka13}. For $\alpha_{x}>\alpha_y$ this zigzag crystal is confined to the xz-plane which prevents it from rotating around the z-axis. 

From eq. \ref{eq:omegay} follows that the static ($\propto\sqrt{q/m}$) and dynamic ($\propto q/m$) parts of the potential scale differently with $q/m$, resulting in a different $\alpha$ for $^{40}$Ca$^{2+}$ compared to singly ionized calcium. The values of $\alpha$  given in this work, always refer to the trapping frequencies of singly charged $^{40}$Ca$^{+}$. To measure the phase transitions of different three-ion crystals the static potentials $\beta$=14.7(2)\,kV/m$^2$ and $\beta_\textnormal{rad}$=37(1)\,kV/m$^2$ are kept constant, the values are determined from the oscillation frequency of a single trapped $^{40}$Ca$^{+}$ ion. The dynamic potential $\gamma$ is varied using a rf-attenuator.

\begin{figure*}
	\centering
		\includegraphics[width=1\textwidth]{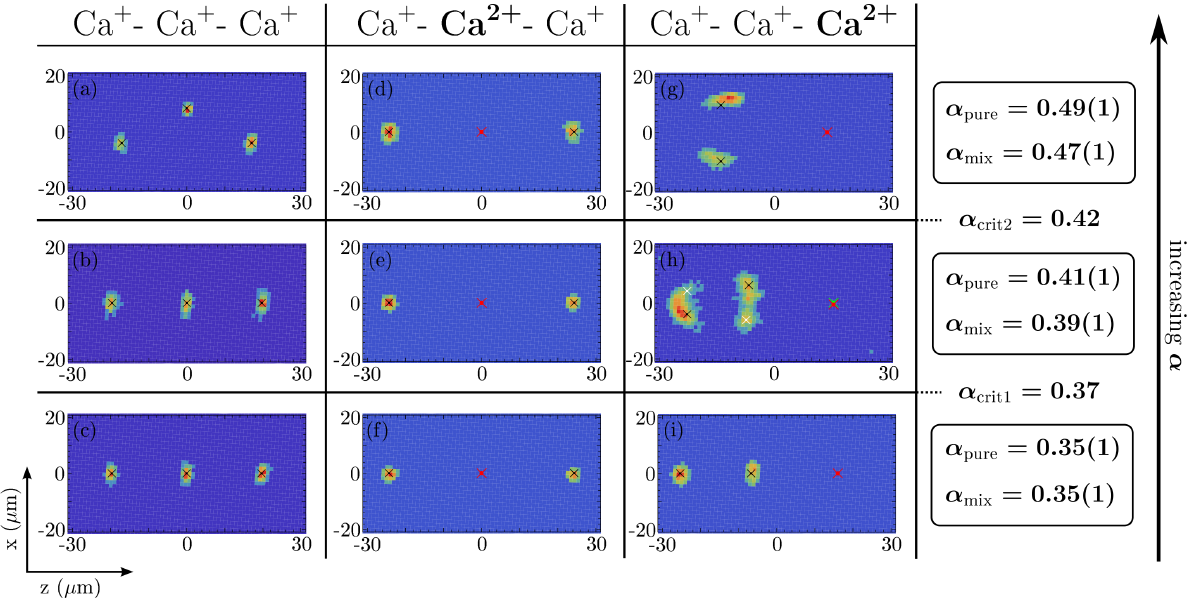} 
	\caption{Fluorescence images (a-i) of three-ion crystals, without $^{40}$Ca$^{2+}$ ion and with $^{40}$Ca$^{2+}$ ion in central or outer position, for different anisotropie parameters. Simulated positions are indicated as black ($^{40}$Ca$^{+}$) and red ($^{40}$Ca$^{2+}$) crosses. In (h) there are additional crosses in white ($^{40}$Ca$^{+}$) and green ($^{40}$Ca$^{2+}$) for the mirrored configuration. The reason for slightly different $\alpha$ for crystals with ($\alpha_{mix}$) and without $^{40}$Ca$^{2+}$ ($\alpha_{pure}$) is a drift of the trapping potential between these measurements.
}
	\label{fig:ZickZackDI-SI}
\end{figure*}

Measured and simulated configurations of the Coulomb crystal are shown in Fig.\ \ref{fig:ZickZackDI-SI}. In the case of a mixed ion crystal $\alpha_\textnormal{crit}$ depends on the position of the $^{40}$Ca$^{2+}$ ion. We observe three different regimes: For $\alpha < \alpha_\textnormal{crit1}$ all crystals are linear strings, for $\alpha_\textnormal{crit1} < \alpha < \alpha_\textnormal{crit2}$ only the crystal with $^{40}$Ca$^{2+}$ in outer position is in zigzag configuration while the others are still linear and for $\alpha > \alpha_\textnormal{crit2}$ the crystal with a central $^{40}$Ca$^{2+}$ ion is the last one remaining in the linear configuration. We could not observe a zigzag crystal with $^{40}$Ca$^{2+}$ in central position at all, instead the $^{40}$Ca$^{2+}$ always moves to the outside for $\alpha \gg \alpha_\textnormal{crit2}$. The theoretical predicted critical anisotropy parameter for the three-ion $^{40}$Ca$^{+}$ crystal $\alpha_\textnormal{crit2}=0.42$ and the $^{40}$Ca$^{2+}$ at the outer position $\alpha_\textnormal{crit1}=0.37$ fits well with the experiment. 

In order to compare observed ion positions of the three-ion crystal with the simulation, a coordinate conversion is applied to the pictures. The line of sight of the imaging system and the crystal plane normal are under 45$^{\circ}$, so the scaling in x- and z-direction is stretched by a factor of $\sqrt{2}$. 	Additionally a deviation of 3$^{\circ}$ between the crystals center line and the z-direction, due to misalignment of the imaging system, has been taken into account.

After this transformation, the deviation between simulated and measured positions is below 1\,$\mu$m for all pictures except for Fig.\ \ref{fig:ZickZackDI-SI}g with a deviation of 3.7\,$\mu$m and Fig.\ \ref{fig:ZickZackDI-SI}h where a fit was not possible. Such deviation may be attributed to a drift of the trapping potential which occurs between calibration and measurement. The additional deviation in Fig.\ \ref{fig:ZickZackDI-SI}g can be attributed to a small error in stray field compensation so that the trapping potential minimum is shifted for $^{40}$Ca$^{2+}$ compared to $^{40}$Ca$^{+}$.

Close to the critical trap anisotropy, when $\alpha$ is barly bigger then $\alpha_\textnormal{crit}$, the potential barrier between mirrored degenerate configurations is small enough to enable the ion to change configurations many times during camera exposure time under Doppler cooling conditions\,\cite{Reis02}. This results in a broadening of the detected peak, as seen in of Fig.\ \ref{fig:ZickZackDI-SI}h and prevents exact determination of ion positions.

\section{Local modes in linear crystal with $^{40}$Ca$^{2+}$ impurity}
\label{localmode}

\begin{figure}[htbp]
	\centering
		\includegraphics[width=0.5\textwidth]{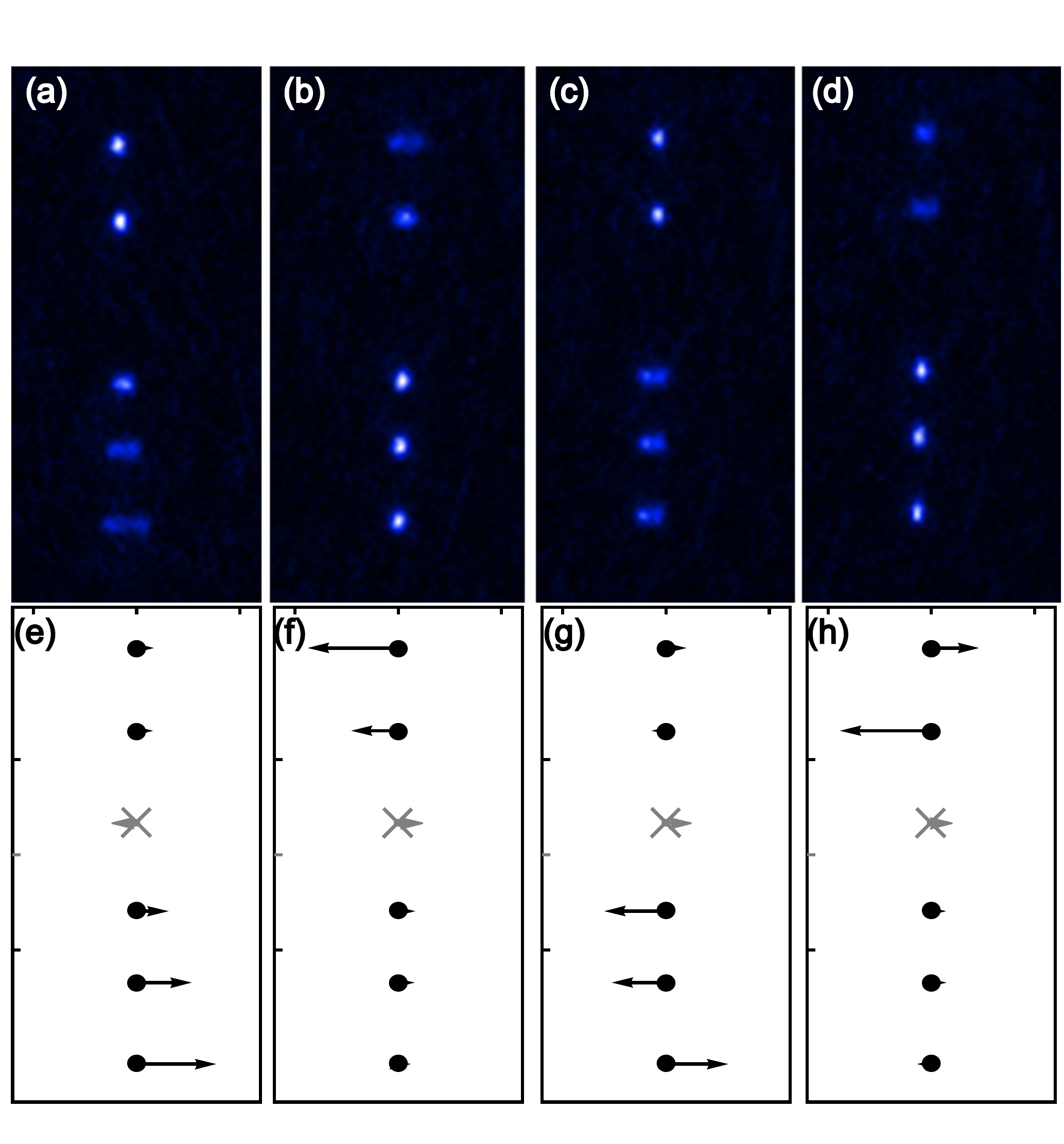}
	\caption{Upper part: Fluorescence images of excited local modes (a-d) in a linear string with $^{40}$Ca$^{2+}$ impurity. The local excitation of ions is observed from the broadening of the ion image in radial direction. The oscillation amplitudes of the ions fit well to the simulated eigenvectors of the modes (e-f).}
	\label{fig:LocalMode}
\end{figure}

A long ion chain can be divided into separately oscillating parts by impurity ions. Here we use a $^{40}$Ca$^{2+}$ ion which is subjected to a much stiffer radial confinement in the trap for the tailoring of radial modes. We prepare a string of five $^{40}$Ca$^{+}$ and one $^{40}$Ca$^{2+}$ positioned as the third ion in the chain and excite the radial modes. We find that for each mode the oscillation is almost completely restricted to either the upper or the lower ions (see Fig.\ \ref{fig:LocalMode}). 

The measured frequencies and eigenvectors are in good agreement with simulations. The eigenfrequencies of the modes are 2$\pi \times$468\,kHz (Fig.\ \ref{fig:LocalMode}a), 2$\pi \times$463\,kHz (Fig.\ \ref{fig:LocalMode}b), 2$\pi \times$423\,kHz (Fig.\ \ref{fig:LocalMode}c) and 2$\pi \times$403\,kHz (Fig.\ \ref{fig:LocalMode}a) for trap frequencies of $\omega_z$=2$\pi \times$119\,kHz, $\omega_x$=2$\pi \times$480\,kHz and $\omega_y$=2$\pi \times$630\,kHz for a single $^{40}$Ca$^{+}$ ion. In addition to the measured modes shown in Fig.\ \ref{fig:LocalMode} there is a mode at 2$\pi \times$1006\,kHz where mainly the $^{40}$Ca$^{2+}$ oscillates and a mode at 2$\pi \times$352\,kHz where the oscillation is localized to the lower subcrystal. Simulations show that for the modes at 2$\pi \times$423\,kHz and at 2$\pi \times$403\,kHz the oscillation amplitude on different sides of the $^{40}$Ca$^{2+}$ differs by a factor of $\approx$\,30 while ions on the same sub-crystal have similar oscillation strength. The results also show that the mode density for the sub-crystals is much lower than for a string of six $^{40}$Ca$^{+}$ ions trapped in the same potential. The minimal frequency separation for modes on the same sub-crystal is 2$\pi \times$45\,kHz compared to 2$\pi \times$15\,kHz for a pure $^{40}$Ca$^{+}$ crystal.

\section{Conclusion and outlook}

We measured the mode-structure of mixed coulomb crystals with three ions. The resulting eigenfrequencies are in good agreement with values from numerical simulations.
We observe the configuration of these crystals the transition from linear to zigzag configuration depending on the position of the doubly-ionized ion. The positions of the $^{40}$Ca$^{+}$ ions agree to 1\,$\mu$m with values from numerical simulations. These results confirm the proposed zigzag transition using Rydberg excitation in a three ion crystal\,\cite{Li12a}. 

We observe localized modes created by dividing longer ion chains with doubly-charged ions. In future experiments such localized modes could possibly be exploited to implement parallel gate execution on different parts of the ion string\,\cite{Li12}.

An exciting perspective for future experiments is the possibility to use coherently excited Rydberg states. The Rydberg excitation from the $3^2$D$_{5/2}$ state, together with the single ion addressing of the $4^2$S$_{1/2}\rightarrow3^2$D$_{5/2}$ transition, may allow for a full control over the radial mode structure. In future, one may extend the mode shaping to two-dimensional ion crystals with an even more complex mode structure\,\cite{Kaufmann12}. 

We thank Weibin Li, Alexander Glaetzle, Rejish Nath, Igor Lesanovsky and Peter Zoller for helpfull discussions. We acknowledge financial support by the EU Chist-Era Project R-Ion and by the BMBF.

%
%

%


\begin{thebibliography}{}

\bibitem{Blatt08}
R. Blatt, D. J. Wineland, Nature \textbf{453,} (2008) 1008.

\bibitem{benhelm08}
J. Benhelm, G. Kirchmair, C. F. Roos, R. Blatt, Nature Phys. \textbf{4,} (2008) 463.

\bibitem{Home09}
J. P. Home, D. Hannecke, J. D. Jost, J. M. Amini, D. Leibfried, D. J. Wineland, Science \textbf{325,} (2009) 1227. 

\bibitem{Ri04}
M. Riebe, H. H\"affner, C. F. Roos, W. H\"ansel, J. Benhelm, G. P. T. Lancaster, T. W. K\"orber, C. Becher, F. Schmidt-Kaler, D. F. V. James, R. Blatt, Nature \textbf{429,} (2004) 734.

\bibitem{La11}
B. P. Lanyon, C. Hempel, D. Nigg, M. M\"uller, R. Gerritsma, F. Z\"ahringer, P. Schindler, J. T. Barreiro, M. Rambach, G. Kirchmair, M. Hennrich, P. Zoller, R. Blatt, C. F. Roos, Science \textbf{334,} (2011) 57-61.

\bibitem{Mo11}
T. Monz, P. Schindler, J. T. Barreiro, M. Chwalla, D. Nigg, W. A. Coish, M. Harlander, W. H\"ansel, M. Hennrich, R. Blatt, Phys. Rev. Lett. \textbf{106,} (2011) 130506.

\bibitem{Lin09}
G.D. Lin, S.L. Zhu, R. Islam, K. Kim, M.S. Chang, S. Korenblit, C. Monroe, L.M. Duan, EPL \textbf{86,} (2009) 60004.

\bibitem{Ga03}
J. J. Garcia-Ripoll, P. Zoller, J. I. Cirac, Phys. Rev. Lett. \textbf{91,} (2003) 157901.

\bibitem{Du04}
L. M. Duan, Phys. Rev. Lett. \textbf{93,} (2004) 100502.

\bibitem{Li12}
W. Li, A.W. Glaetzle, R. Nath, I. Lesanovsky, Phys. Rev. A \textbf{87,} (2013) 052304.

\bibitem{Schmidt-Kaler11}
F. Schmidt-Kaler, T. Feldker, D. Kolbe, J. Walz, M. M\"uller, P. Zoller, W. Li, I. Lesanovsky, New J. of Phys. \textbf{13,} (2011) 075014.

\bibitem{Mueller08}
M. M\"uller, L. Liang, I. Lesanovsky, P. Zoller, New J. of Phys. \textbf{10,} (2008) 093009.

\bibitem{Li12a}
W. Li, I. Lesanovsky, Phys. Rev. Lett. \textbf{108,} (2012) 023003.

\bibitem{Ulm13}
S. Ulm, J. Ro{\ss}nagel, G. Jacob, C. Deg\"unther, S. T. Dawkins, U. G. Poschinger, R. Nigmatullin, A. Retzker, M. B. Plenio, F. Schmidt-Kaler, K.Singer, arXiv:13025343v1 (2013).

\bibitem{Pyka13}
K. Pyka, J. Keller, H. L. Partner, R. Nigmatullin, T. Burgmeister, D. M. Meier, K. Kuhlmann, A. Retzker, M. B. Plenio, W. H. Zurek, A. Campo, T. E. Mehlst�ubler, arXiv:1211.7005v1 (2012)

\bibitem{Ba12}
J. D. Baltrusch, C. Cormick, G. Morigi, Phys. Rev. A \textbf{86,} (2012) 032104.

\bibitem{Kolbe12}
D. Kolbe, M. Scheid, J. Walz, Phys. Rev. Lett. \textbf{109,} (2012) 063901.

\bibitem{Ko12}
D. Kolbe, M. Scheid, J. Walz, Appl. Phys. B (2012) 10.1007/s00340-013-5510-6.

\bibitem{Kw07}
T. Kwapien, U. Eichmann, W. Sandner, Phys. Rev. A \textbf{75,} (2007) 063418.

\bibitem{Akerman10}
N. Akerman, S. Kotler, Y. Glickman, Y. Dallal, A. Keselman, R. Ozeri, Phys. Rev. A, Rapid Communications \textbf{82,} (2010) 061402.

\bibitem{Leibfried03}
D. Leibfried, R. Blatt, C. Monroe, D. Wineland, Rev. Mod. Phys. \textbf{75,} (2003) 281.

\bibitem{Reis02}
D. Rei{\ss}, K. Abich, W. Neuhauser, Ch. Wunderlich, P. E. Toschek, Phys. Rev. A \textbf{65,} (2002) 053401 

\bibitem{Enzer00}
D.G. Enzer, M.M. Schauer, J.J. Gomez, M.S. Gulley, M.H. Holzscheiter, P.G. Kwiat, S.K. Lamoreaux, C.G. Peterson, V.D. Sandberg, D. Tulpa, A.G. White, R.J. Hughes, D.F.V. James, Phys. Rev. Lett. \textbf{85,} (2000) 2466.

\bibitem{Kaufmann12}
H. Kaufmann, S. Ulm, G. Jacob, U. Poschinger, H. Landa, A. Retzker, M.B. Plenio, F. Schmidt-Kaler, Phys. Rev. Lett. \textbf{109,} (2012) 263003.


%
%
\end{thebibliography}


\end{document}